\documentclass[10pt, a4paper, twocolumn]{article}

\usepackage[colorlinks=true, citecolor=blue!50!black]{hyperref}

\usepackage[english]{babel} 

\usepackage{microtype} 

\usepackage{amsmath,amsfonts,amsthm} 

\usepackage[svgnames,x11names,table]{xcolor} 

\usepackage[hang, small, labelfont=bf, up, textfont=it]{caption} 

\usepackage{booktabs} 

\usepackage{lastpage} 

\usepackage{graphicx} 

\usepackage{enumitem} 
\setlist{noitemsep} 

\usepackage{sectsty} 
\allsectionsfont{\usefont{OT1}{phv}{b}{n}} 

\usepackage{geometry} 

\usepackage[T1]{fontenc} 
\usepackage[utf8]{inputenc} 

\usepackage{XCharter} 

\usepackage{fancyhdr} 
\fancypagestyle{firstpage}{ 
	\fancyhf{}
}

\newcommand{\authorstyle}[1]{{\usefont{OT1}{phv}{b}{n}\color{Black}#1}} 

\newcommand{\institution}[1]{{\footnotesize\usefont{OT1}{phv}{m}{sl}\color{Black}#1}} 

\usepackage{titling} 

\newcommand{\HorRule}{\color{DarkGoldenrod}\rule{\linewidth}{1pt}} 

\pretitle{
	\vspace{-50pt} 
	\HorRule\vspace{10pt} 
	\fontsize{20}{28}\usefont{OT1}{phv}{b}{n}\selectfont 
	\color{DarkRed} 
}

\posttitle{\par\vskip 15pt} 

\preauthor{} 

\postauthor{ 
	\vspace{10pt} 
	\par\HorRule 
	\vspace{8pt} 

}

\usepackage{lettrine} 
\usepackage{fix-cm}	

\usepackage{xstring} 

\usepackage[english]{babel}
\usepackage{amsmath}
\usepackage{bbold}
\usepackage{graphicx}
\usepackage{mdframed}
\usepackage{amssymb}
\usepackage{cite}
\usepackage{caption}
\usepackage[table]{xcolor}
\usepackage{multicol}

\newcommand{\ttilde}{\widetilde}

\title{Degradation rate uniformity determines success of oscillations in repressive feedback regulatory networks}
\author{Page, Karen M. \and Perez-Carrasco, Ruben}

\usepackage{newfloat}
\DeclareFloatingEnvironment[fileext=frm,placement={!ht},name=Frame]{myfloat}
\author{\authorstyle{Karen M. Page\textsuperscript{1} \& Ruben Perez-Carrasco\textsuperscript{1}}\\
\institution{\textsuperscript{1} Department of Mathematics, University College London, Gower Street, WC1E 6BT, London, UK\\
}}

\begin{document}
\maketitle
\date{\today}
\section*{Summary}
\begin{mdframed}[backgroundcolor=black!20,linewidth=0,leftmargin=-0.5cm,rightmargin=-0.3cm]
\textbf{Ring oscillators are biochemical circuits consisting of a ring of interactions capable of sustained oscillations. The non-linear interactions between genes hinder the analytical insight into their function, usually requiring computational exploration. Here we show that, despite the apparent complexity, the stability of the unique steady state in an incoherent feedback ring depends only on the degradation rates and a single parameter summarizing the feedback of the circuit. Concretely, we show that the range of regulatory parameters that yield oscillatory behaviour, is maximized when the degradation rates are equal. Strikingly, this results holds independently of the regulatory functions used or number of genes. We also derive properties of the oscillations as a function of the degradation rates and number of nodes forming the ring. Finally, we explore the role of mRNA dynamics by applying the generic results to the specific case with two naturally different degradation time scales.}
\end{mdframed}
\thispagestyle{firstpage} 

\section*{Introduction}

Genetic regulatory networks (GRNs), consisting of the interactions between a set of genes, are core to the regulation of the temporal genetic expression profiles required for various cellular processes, ranging from cell fate determination during embryogenesis to cellular homesotasis \cite{Davidson2005,Levine2005,Panovska-Griffiths2013,Olson2006,Sauka-Spengler2008,Dequant2006}. GRNs are capable of many dynamical functions, including oscillatory gene expression \cite{Monk2003}, as has been observed in somitogenesis \cite{Hirata2002}, circadian clocks \cite{Reddy2014}, the activity of the p53 tumor suppressor \cite{Bar-Or2000,Michael2003}) or the nuclear factor $\kappa$B localization \cite{Hoffmann2002}.

Due to their range of utilities, different oscillatory gene regulatory circuits have been synthetically engineered \cite{Purcell2010}. In particular, lot of attention has been focused on the engineering of ring oscillators consisting of a set of genes interacting with each other sequentially and forming a repressive feedback loop. This work was initiated by the synthesis of the 3-gene repressilator \cite{Elowitz2000}, that has been further refined to improve its oscillation properties (e.g. \cite{Stricker2008,Niederholtmeyer2015}). Consequently, the theoretical and numerical analysis of the working of ring oscillators has also received substantial attention. Such work was pioneered by Fraser and Tiwari \cite{Fraser1974} who performed numerical simulations. Subsequent analysis showed that for sufficiently strong repression, oscillations arise due to a Hopf bifurcation, relating the genetic oscillatory behaviour with dynamical systems theory \cite{Smith1987}, which has lead to many different studies delving into dynamical properties of the oscillations  (e.g. \cite{Buse2009, Buse2010, Mueller2006, Garcia-Ojalvo2004,Pigolotti2007})


These analytical and numerical studies of biochemical circuits require insight into a set of simultaneous non-linear feedback interactions between multiple genes usually analyzed as a set of ordinary differential equations (ODEs). Determining the role of different parameters in the solutions to these equations poses enormous analytical complexity that hinders quantitative studies. For this reason, computational and analytical studies are often reduced to tackling relatively small networks, and, even in such cases, to a reduced parameter set or certain simplified regulatory functions. This can constrain the range of application of the results found \cite{Estrada2016}. Even in the case of the repressilator, the dynamical complexity can be huge \cite{Potapov2015} and restrictive assumptions within the quantitative model again become unavoidable. This highlights the necessity to develop tools capable of understanding dynamical properties of the system independently of the regulatory functions used. 


A useful assumption, present in the vast majority of studies, is that the degradation rates of proteins are identical for different genes. However, due to the high span of protein structures and mechanisms controlling degradation rates, such as ubiquitination \cite{Bachmair1986,Dice1987}, the turnover rate can range orders of magnitude in the proteome of a single system \cite{Belle2006,Christiano2014}. Since oscillations in a network are generated by an ongoing imbalance between the production and degradation of the different species, it is expected that degradation rates play a determinant role in the behavior of oscillatory circuits. Particuarly, simulations of a repressilator model showed that oscillations are favoured for comparable values of the degradation of the protein and mRNA \cite{Elowitz2000}, and, more generally, a certain level of symmetry around the ring \cite{Tuttle2005}. Nevertheless, there is no analytical study that gives insight into the role of degradation rates for general ring oscillators independent of the regulatory functions used.

To gain insight into the role of degradation rates in oscillatory networks, dynamical system theory and bifurcation theory have proven to be essential tools. These allow us to categorise different possible dynamical responses of oscillatory networks \cite{Smith1987,Strelkowa2010,Parmar2015,Monk2003}. Using bifurcation theory we aim to obtain information on the role of degradation rates in oscillatory networks, making these results as general as possible and using minimal details of the regulatory functions. Specifically, we show how relevant information on the interactions between different genes can be captured with a single parameter. We show how this parameter controls the appearance of oscillations through a Hopf bifurcation. First we develop our methodology for the repressilator, expanding the theory in the following sections to negative feedback rings oscillators with an arbitrary number of species. Finally we study the case in which the species are categorized as mRNAs and proteins, which have distinct degradation rates, giving insight in the role of mRNA dynamics in the performance of ring oscillators.

\section*{Results}

\subsection*{Three-gene repressilator}

\begin{figure*}
\centering
\includegraphics[width = 2.0 \columnwidth]{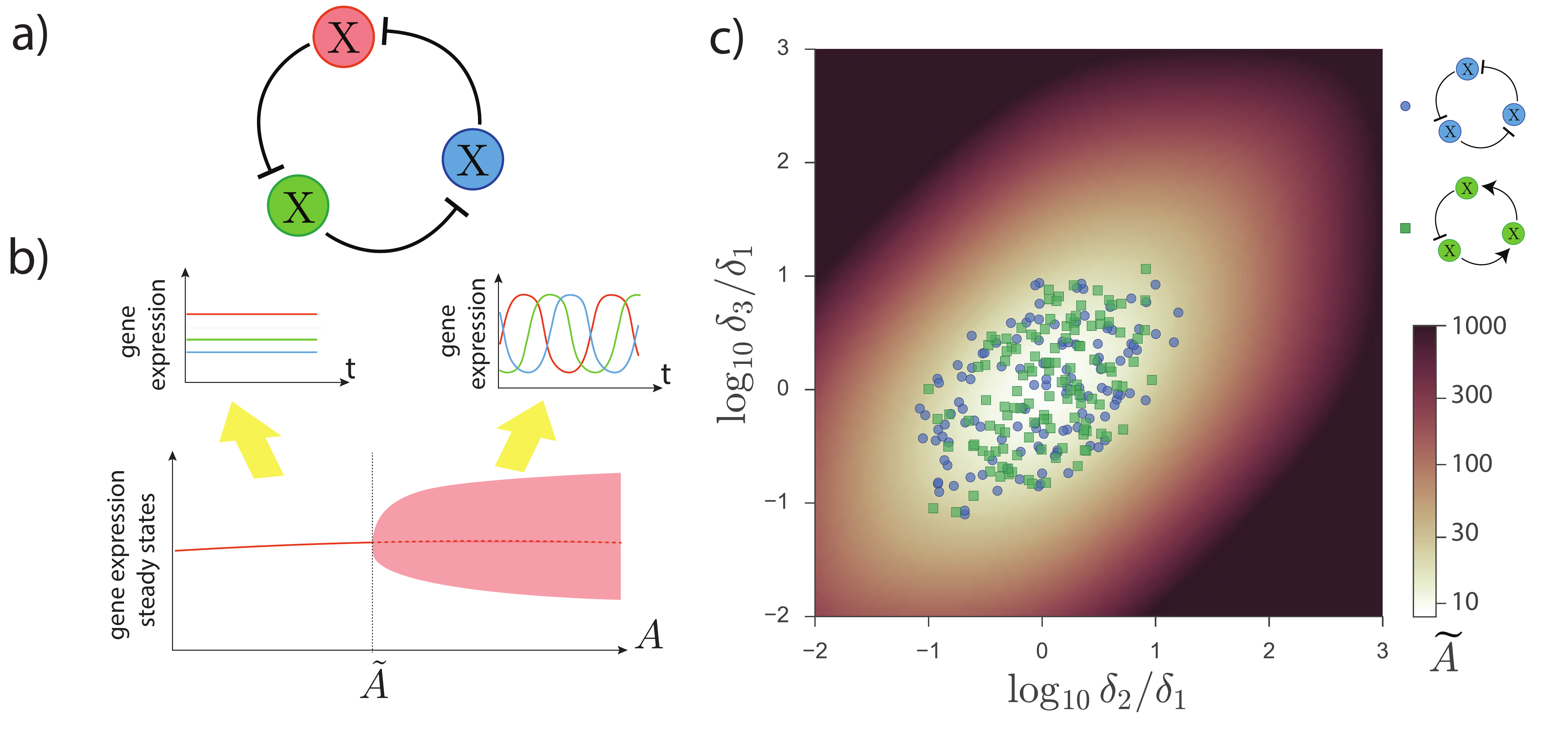}
\caption{\label{fig.rep} \label{fig.sample3d}
Oscillatory behaviour of the repressilator a) Schematic of the repressilator b) Bifurcation diagram schematic shows how the oscillations appear and disappear through a Hopf bifurcation depending on the magnitude $A$ that summarizes the negative feedback strength of the circuit. c) Degradation rate relationship of repressilator networks showing oscillations from a random screening (\emph{squares and circles}). Different symbols stand for the screening of the repressilator (\emph{blue circles}), and the repression ring with only one repression and two activations (\emph{green squares}). Random repressilator networks were generated by sampling random parametrizations of $f_i$ and sampling relative degradation rates covering the whole square plotted ($\delta_2/\delta_1$ and $\delta_3/\delta_1$ between $[10^{-2},10^3]$, $\delta_1=1$ in all simulations). Results are compared with the  value of $\ttilde A$ for different degradation rates (contour plot from Equation \ref{eq.Acond3d}). Random interaction functions were generated using the thermodynamic function $f_i(x)=a_i\left(1+\rho^R_i(1+x/k_i)^h\right)^{-1}$ for the repressions and  $f_i(x)=a_i\left(1+\rho^A_i[(1+x/k_i)/(1+l_ix/k_i)]^h\right)^{-1}$ for the activations with $h=3$. Random parameters were sampled logarithmically from the intervals $k_i: [10^{-9},10^{-5}]$, $a_i : [10^{-4},10^4]$, $\rho^R_i :  [10^{-4},10^4]$, $\rho^A_i :  [10^3,10^{11}]$, $l_i :  [10,10^5]$.}
\end{figure*}

The classic general form of the repressilator consists of three genes repressing each other sequentially \cite{Elowitz2000} (Fig. \ref{fig.rep}a). In the simple case in which mRNA dynamics are considered fast compared with protein dynamics, the dynamical evolution of the system can be described as a set of ODEs

\begin{eqnarray}
\dot{x}_1 & = & \delta_1(f_1(x_3)- x_1) \nonumber \\
\dot{x}_2 & = & \delta_2(f_2(x_1)- x_2) \nonumber \\
\dot{x}_3 & = & \delta_3(f_3(x_2) -x_3) \nonumber, 
\end{eqnarray}
where $f_1, f_2$ and $f_3$ describe the repressive interactions between genes and are therefore decreasing, positive functions. $f_i$ can be thought of as the maximal expression level of gene $i$ multiplied by the probability that its repressor is inactive. At any given steady state $(x_1^*,x_2^*,x_3^*)$ given by $\dot x_1=\dot x_2 =\dot x_3 = 0$, the repressilator follows the relationship $x_3^* = f_3(f_2(f_1(x_3^*))) \equiv F(x_3^*)$, where the function $F$ captures the overall negative feedback. Since $F(x)$ is a decreasing, positive function, there is a unique possible value of $x_3^*$, which yields unique values $x_1^*=f_1(x_3^*)$ and $x_2^*=f_2(x_1^*)$. The stability of the protein levels dictated by this steady state, can be computed through the eigenvalues of its Jacobian matrix
\begin{equation}
J = \left( \begin{array}{ccc} -\delta_1 & 0 & \delta_1f_1'(x_3^*) \\ \delta_2f_2'(x_1^*) & -\delta_2 & 0 \\ 0 & \delta_3f_3'(x_2^*) & -\delta_3 \end{array} \right).
\end{equation} 

These eigenvalues  $\lambda$ satisfy the characteristic equation
\begin{equation}
\label{eq.charac3}
(\lambda+\delta_1)(\lambda+\delta_2)(\lambda+\delta_3) +A \delta_1\delta_2\delta_3= 0,
\end{equation}
where $A \equiv -f'_1(x_3^*)f'_2(x_1^*)f'_3(x_2^*) = -F'(x_3^*)$ is the modulus of the slope of the composite repression function at the steady state. Interestingly, the parameter A contains all the details of the interactions of the network necessary to solve the characteristic equation (\ref{eq.charac3}). This means that the eigenvalues of the characteristic equation and so the stability of the steady protein levels will depend only on $A$ and on the degradation rates. This allows us to perform the stability analysis without any further information on the explicit form of the repressive interactions. Concretely,  since $F(x)$ is a monotonically decreasing function ($A > 0$) the product of the eigenvalues of $J$ will always be negative,
\begin{equation}
 \lambda_1\lambda_2 \lambda_3 = \det J =  -\delta_1\delta_2\delta_3(1 + A)<0.
\label{eq.ineqlll}
\end{equation} 
Therefore, the repressive ring forbids any eigenvalue to be zero. As a result, a change in stability of the steady state can only occur through a Hopf bifurcation, in which a pair of complex conjugate eigenvalues crosses the imaginary axis. We write this pair $\ttilde\lambda_2=i\alpha$ and $\widetilde\lambda_3 = -i\alpha$, where $\alpha$ is the angular velocity of the sustained oscillations that appear at the Hopf bifurcation. Following Equation(\ref{eq.ineqlll}), the other eigenvalue $\lambda_1$ must be real and negative, everywhere, and in particular at the Hopf bifurcation ($\ttilde\lambda_1<0$). 

Introducing the purely imaginary eigenvalues $\ttilde\lambda_2$ and $\ttilde\lambda_3$ in the characteristic equation (\ref{eq.charac3}), expressions for $\alpha$ and $\tilde A$ (value of $A$ at the bifurcation) are obtained that only depend on the degradation rates,
\begin{equation}
\label{eq.alpha3d}
\alpha= \sqrt{\delta_1\delta_2+\delta_2\delta_3+\delta_3\delta_1}.
\end{equation}
 \begin{equation}
 \label{eq.Acond3d}
\ttilde A = \frac{(\delta_1\delta_2+\delta_2\delta_3+\delta_3\delta_1)(\delta_1+\delta_2+\delta_3)}{\delta_1\delta_2\delta_3} -1.
\end{equation}
Since the value of $\ttilde A$ is unique, the repressilator has a single Hopf bifurcation with gene expression $(\ttilde{x}_1,\ttilde{x}_2,\ttilde{x}_3)$. Concretely, at the lowest value of $A$ ($A=0$), the eigenvalues of the Jacobian are all negative $(\lambda_i=-\delta_i,~ i=\lbrace1,2,3\rbrace)$, and the steady state is stable. Since there is a change in the stability of the steady state at $\tilde A$, the steady state is stable for $A<\ttilde A$ and unstable (with the appearance of a stable oscillatory orbit) for $A>\ttilde A$. Thus, the smaller the value of $\ttilde A$, the easier it is to find oscillations in the system (Fig. \ref{fig.rep}b). Strikingly, the value of  $\ttilde A$ just depends on the degradation rates (see Equation (\ref{eq.Acond3d})) and is minimized when they are equal, $\delta_1=\delta_2=\delta_3$, giving $\min\{\ttilde A\}\equiv\ttilde A_m=8$.  Therefore the closer the degradation rates are to being equal, the less strict is the condition on the network parameters through $A$ in order for the system to oscillate. This has been tested computationally by generating random repressilator networks, showing that knowledge of the value of $\ttilde A_m$ derived from the degradation rates gives a prediction of the propensity for oscillations of the repressilator network (Fig. \ref{fig.sample3d}c). 

It is interesting to note that the three-gene repressilator analysis extends straightforwardly to the negative feedback loop case consisting of two activations and one repression. In this case, the composite function $F$ is again a positive decreasing function which is the only requirement in our analysis, yielding exactly the same results (Fig. \ref{fig.sample3d}c). 
\subsection*{N-component negative feedback ring}

The reduced three-gene scenario considered above does not include intermediate mRNA dynamics or other intermediate regulatory steps. Additionally, it is not straightforward to apply the results to repressive rings with a gene number greater than three, such as the artificial circuits created by \cite{Niederholtmeyer2015}. In order to analyze these systems, we can extend the repressilator by considering the general case of N biochemical species as an N-dimensional monotone cyclic feedback system \cite{Mallet-Paret1990}:
\begin{equation}
\label{eq.Ndimrep}
\dot x_n=\delta_n (f_n(x_{n-1})-x_n), \mbox{for }n=1,...N,
\end{equation} 
where $x_0 \equiv x_N$. To ensure that the network presents a negative feedback loop, it must contain an odd number of repressions $N_R$, \emph{i.e.} $N_R$ of the functions $f_n$ are monotonic decreasing positive functions. In addition there are $N_I= N-N_R$ activations, $i.e.$ $N_I$ of the functions  $f_n$ are monotonic increasing positive functions. In order to extend the result to $N$ components, we will follow a derivation equivalent to that of the repressilator. In this case the steady state is located at $x^*_N=f_N(f_{N-1}(....f_1(x^*_N)))\equiv F(x^*_N)$. As in the three-gene repressilator, $F(x)$ is a positive monotonically decreasing function and so there is a single value for $x^*_N$ and hence a unique steady state $x^*_1=f_1(x^*_N)$, $x^*_2=f_2(x^*_1)$, ..., $x^*_{N-1}=f_{N-1}(x^*_{N-2})$.

Since the N-component repressive ring cannot show chaotic behaviour (see SI), when the steady state is unstable, it will not be able to attract trajectories, and orbits will converge to a limit cycle where all the biochemical species will oscillate in time. As in the repressilator, this allows us to study the oscillatory properties of the GRN through its Jacobian $J$ at the steady state $x^*$
\begin{equation}
J = \left( \begin{array}{ccccc} -\delta_1 & 0& ...& 0 & \delta_1f_1'(x^*_N) \\ \delta_2f_2'(x^*_1) & -\delta_2 & 0& ... & 0 \\  \vdots & & \ddots& \ddots& \vdots \\ 0 & ... & 0 & \delta_Nf_N'(x^*_{N-1}) & -\delta_N \end{array} \right).
\end{equation}
The corresponding characteristic equation is given by
\begin{equation}
\prod_{n=1}^N (\lambda+\delta_n) - \prod_{n=1}^N \delta_n f_n'(x^*_{n-1})=0 . \nonumber
\end{equation}
As in the three-gene case, the chain rule for differentiation gives us $-F'(x^*_N)=-\prod_{n=1}^N f_n'(x^*_{n-1})\equiv A > 0$ and hence the characteristic equation can be written as 
\begin{equation}
\label{eq.characteristic}
\prod_{n=1}^N \left(1+\frac{\lambda}{\delta_n} \right) + A = 0.
\end{equation}
Therefore, as with the classic repressilator, despite all the potential complexity in the repression functions of the network, the stability of the unique steady state only depends on the degradation rates and the parameter $A$, which gathers information on the global negative feedback loop, as the modulus of the slope of the composite repression function of a gene on itself at the steady state. The product of the eigenvalues of $J$ also follows the same pattern as Equation \ref{eq.ineqlll},
\begin{equation}
\label{eq.detJN}
\prod_{n=1}^N{\lambda_n} = \det{J} = (-1)^N(1+A)\prod_{n=1}^N \delta_n \neq 0,
\end{equation}
forbidding a zero eigenvalue of $J$, so the steady state can only lose stability via a Hopf bifurcation (see SI). At $A=0$, the Jacobian matrix has roots $\lambda=-\delta_n$ for $n=1,..,N$, and is therefore a stable node. Increasing $A$ away from zero, oscillations arising through a Hopf bifurcation will appear at the smallest value $A=\ttilde A$ at which $J$ has a pair of imaginary eigenvalues. Letting the pair of eigenvalues be $\pm i \alpha$ with the angular velocity $\alpha>0$ and introducing it in Equation \ref{eq.detJN} we can again derive relationships for $\alpha$ and $\tilde A$ that only depend on the degradation rates (see SI), 
\begin{equation}
\label{eq.tildeAmoduli}
\prod_{n=1}^N \left( 1 + \frac{\alpha^2}{\delta_n^2} \right) = {\ttilde A}^2.
\end{equation}
\begin{equation}
\label{eq.implicit}
\sum_{n=1}^N \tan^{-1} (\alpha/\delta_n) = \pi.
\end{equation}

Note that in contrast to the three-gene case, there is no closed form expression for the angular velocity $\alpha$ and $\ttilde A$ as a function of the degradation rates comparable to Equations \ref{eq.alpha3d} and \ref{eq.Acond3d}. Instead we have the Implicit Equation (Equation \ref{eq.implicit}) that returns the value of the angular velocity $\alpha$ for a certain set of values $\delta_n$ and Equation \ref{eq.tildeAmoduli} that returns the value of $\ttilde A$, once $\alpha$ is known. 

As in the previous section, we are interested in the degradation rates for which  $\ttilde A$ is minimized, since this will maximize the parameter region for which there will be oscillations.  For this purpose we can work with the arguments  $\theta_n \equiv \tan^{-1}( \alpha/\delta_n)$ varying independently in the domain $[0,\pi/2)$ subject to the constraint that they sum to $\pi$. In this representation $\ttilde A=\prod_{n=1}^N \sec \theta_n$ from Equation \ref{eq.tildeAmoduli}. To find its minimum value, we minimize $\ln \ttilde A$ introducing the Implicit Equation (\ref{eq.implicit}) constrain ($\sum_{n=1}^N \theta_n = \pi$) using the Lagrange multiplier $\mu$,
\begin{equation}
\frac{\partial \ln \ttilde A}{\partial \theta_n} - \mu \frac{\partial \sum_{l=1}^N \theta_l }{\partial \theta_n} = 0,\mbox{ }n=1,...,N, \nonumber
\end{equation}
which yields
\begin{equation}
\label{eq.tanmu}
\tan \theta_n = \mu,\quad n=1,...,N.
\end{equation}
Since $\theta_n$ vary in the domain $[0,\pi/2)$, the condition (\ref{eq.tanmu}) is only fulfilled when all $\theta_n$ are the same ($\theta_n = \pi/N$). It is straightforward to check that this stationary point is a minimum since $\ttilde A$ can be made arbitrarily big by choosing $\theta_1\rightarrow \pi/2$. Thus, as in the classic three-gene repressilator, the minimum value of $\ttilde A\equiv\ttilde A_m$ is achieved when all the degradation rates are equal. For this case, an analytical expression for $\ttilde A$ is available  from Equation \ref{eq.tildeAmoduli},
\begin{equation}
\label{eq.minA}
{\ttilde A_m} =\sec^{N} (\pi/N). 
\end{equation}
We can show that $\tilde A_m$ is decreasing in $N$, for $N\ge3$.  Therefore increasing $N$ increases the range of values of $A$ for which we get oscillations. As $N\rightarrow \infty$, the critical value of $A$ tends to $1$, whilst when $N=3$, the prediction $\ttilde A_m = 8$ is recovered.

Additionally, an expression for the angular frequency $\alpha\equiv \alpha_m$ of the small oscillations that arise close to this bifurcation point when all the degradation rates are identical $\delta_n\equiv \delta$ is also available,
\begin{equation}
\label{eq.minalpha}
\alpha_m=\delta \tan (\pi/N).
\end{equation}
Like $\tilde A_m$, the frequency $\alpha_m$ is decreasing in $N$ showing that the more links the feedback loop has, the slower oscillations will get.  For $N=3$, the results from the first section are recovered, predicting an angular frequency is $\delta\sqrt{3}$, so that the time period of oscillations is $2\pi/\sqrt{3} \approx 3.62/\delta$. In the limit $N \rightarrow \infty$, the transmission of information across the feedback gets infinitely slow and the frequency of the oscillations tends to $0$. The slowing down of the oscillations with $N$ is also true for the general case in which the degradation rates are not identical. If we fix $\delta_1,\delta_2,...,\delta_N$ and consider introducing an additional species $x_{N+1}$ in the cycle (keeping the number of repressions odd), then it is clear from the Implicit Equation (Equation \ref{eq.implicit}) that the value of $\alpha$ which satisfies this equation will be lowered. 

\begin{figure*}
\centering
\includegraphics[width=2.0\columnwidth]{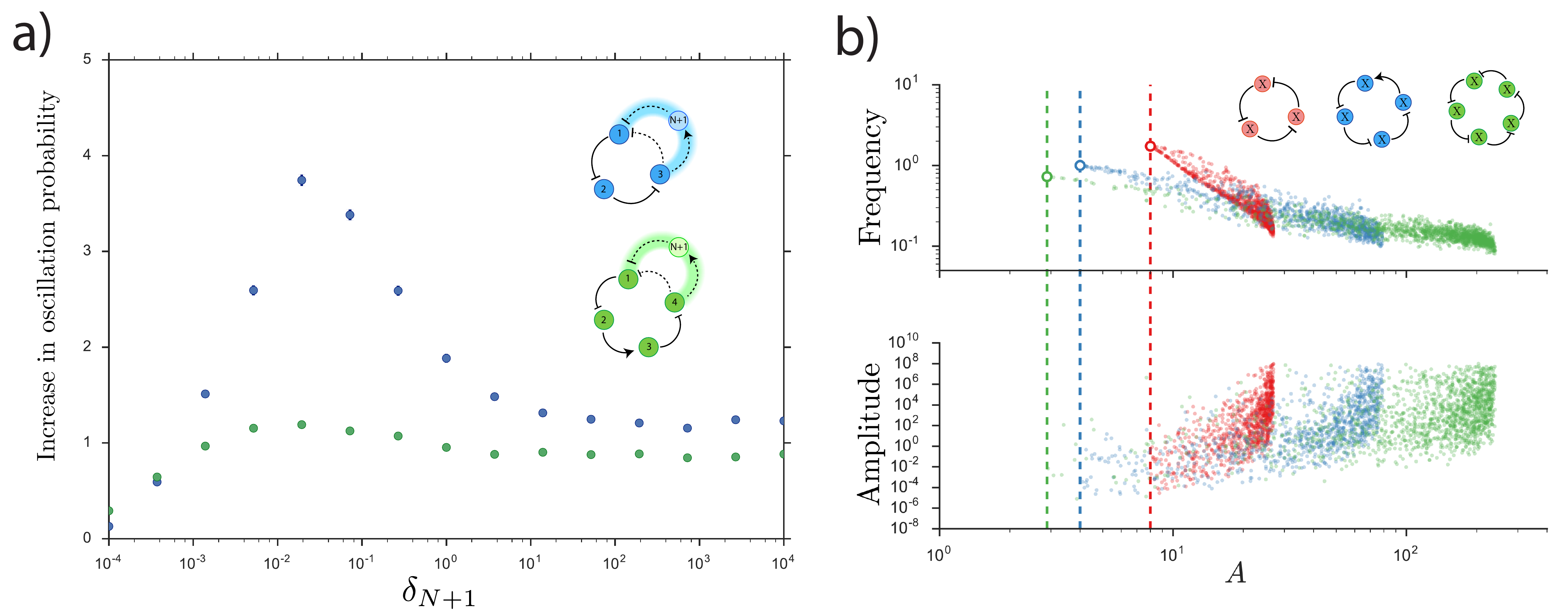}
\caption{\small\label{fig.newlint} \label{fig.freqplotsame}
Behaviour of N-gene oscillators. a) Ratio of the probability of oscillations in an extended ring network with $N+1$ genes and a genetic network of $N$ genes for different values of the degradation rate of the $N+1$th gene. In both cases explored N=3 (blue) and N=5 (green), the degradation of the $N+1$th gene is varied while sampling the other degradation rates from the range $\delta_i=[10^{-3},1]$ and keeping one random gene fixed at $\delta=1$. The other parameters and details of the sampling are the same as in Fig. \ref{fig.sample3d}c. b) Frequency and amplitude of oscillations as a function of the network parameter $A$ for 1000 successfully oscillatory networks from a random screening for different numbers of species with the same degradation rates $\delta_i=1$, $N=3$ (\emph{red}), $N=4$ (\emph{blue}) and $N=5$ (\emph{green}). Dashed lines and rings show the minimum critical value $\tilde A_m$ and angular velocity at that point $\alpha_m$  predicted by Equations \ref{eq.minA} and \ref{eq.minalpha} for $N=3,4,5$. The repression functions are the same as in Fig. \ref{fig.sample3d}c with parameters logarithmically sampled from the intervals $k_i: [10^{-5},10^{-1}]$, $a_i : [10^{-4},10^8]$, $\rho^R_i :  [10^{-4},10^4]$, $\rho^A_i : [10^{-1},10^{11}]$, $l_i : [10^{1},10^{5}]$.}
\end{figure*}

On the contrary, introducing a new species does not necessarily reduce $\ttilde A$. In order to evaluate the effect on $\ttilde A$ of adding a new link it is interesting to note first that in the case that the degradation rate of the new species tends to infinity ($\delta_{N+1}\rightarrow\infty$ in Equations \ref{eq.implicit} and \ref{eq.tildeAmoduli}), the problem is reduced to the case with $N$ species, \emph{i.e.} the new variable will be so fast that will be always in quasi-equilibrium with the previous species. In contrast, introducing an arbirtrarily slow degrading species will completely stop the oscillations. It can be proved (see SI) that there is a range of degradation rates of the new species for which the probability of oscillations is increased. This is supported by numerical simulations (Figure \ref{fig.newlint}a). The relative probability of oscillations does not precisely tend to one as the degradation rate of the added species tends to infinity, because the value of $A$ is also changed by addition of the extra species; simulations in which the added species has $f_{n+1}(x) =x$, the relative probability of oscillations tends to one as $\delta_{N+1} \rightarrow \infty$ (data not shown). The simulations also demonstrate that the increase in probability of oscillations with additional species of intermediate degradation rate becomes weaker the more species there are.

\subsection*{Is the Hopf bifurcation supercritical or subcritical?}

In the development of the argument we assumed that the Hopf bifurcation is supercritical and not subcritical \emph{i.e.} a stable limit cycle arises at the bifurcation point. This is true for all the networks explored numerically in this manuscript, which use thermodynamic regulatory functions. Nevertheless, this is not necessarily true for any repressive functions $f_i(x)$. Mathematically, this requires the computation of the sign of the first Lyapunov coefficient \cite{Kuznetsov2013} at the Hopf bifurcation. In the case where degradation rates and repressive functions are the same for all species (an assumption that is often made, e.g. \cite{Elowitz2000}), progress can be made. For the three-gene repressilator with Hill function repressions it can be proved that the Lyapunov coefficient is negative, \emph{i.e.} there is always a supercritical bifurcation \cite{Buse2009}. In the case of an N-component repressive ring, the first Lyapunov coefficient $\ell_1$ is given by (see SI),

\begin{equation} \label{supercrit}
 \ell_1 = \frac{c^2}{2N\sin(\pi/N)}  \left[ -f'''(\tilde x)+\frac{f''(\tilde x)^2[4c^3+4c^2-13c+2]}{(1+c)(5-4c)} \right],
\end{equation}
where $c=\cos(\pi/N)$. Therefore the sign of $\ell_1$will depend on the ratio $\frac{f'''(\tilde x)}{f''(\tilde x)^2}$ and the number of links. While the Lyapunov coefficient is negative for the thermodynamic regulatory functions chosen in this manuscript and for Hill function repressions, the Lypaunov coefficient is not negative for every possible regulatory function $f$. For example a repressive ring with $f(x)=1/(1+(1+x-x^2+x^3)^h)$ can have positive or negative coefficient depending on the exponent $h$  (see SI and Fig.\ref{fig.lyapunov}). Nevertheless, since trajectories for genetic systems are bounded, the unstable limit cycle must coexist with a stable limit cycle for $A>\ttilde A$, returning a comparable set of results even in the case a subcritical bifurcation occurs. In this case, stable oscillations or evolution towards a steady state concentration will both be possible for values of $A$ slightly lower than $\ttilde A$.

As mentioned, the nature of the Hopf bifurcation depends on $\frac{f'''(\tilde x)}{f''(\tilde x)^2}$. If it is greater than $-2/3$, then the Hopf bifurcation is supercritical for all $N$ for which it exists. If it is less than $-3/2$, then the Hopf bifurcation is subcritical for all $N$ for which it exists. If $-3/2<\frac{f'''(\tilde x)}{f''(\tilde x)^2}<-2/3$, then the Hopf bifurcation is supercritical for sufficiently large $N$ and subcritical for smaller $N$, assuming it exists.

\subsection*{Behaviour away from the Hopf bifurcation}

We have shown that for value of $A$ below the critical value ($A<\ttilde A$), the steady state is stable, and changes stability at $A=\ttilde A$. But, can the stability be recovered for greater values of $A$?, or in other words, is it guaranteed that the oscillations will be stable for all values $A>\ttilde A$? In order to answer this question we can count the maximum number of pairs of eigenvalues crossing the imaginary axis as the possible solutions of the Implicit Equation and also the number of pairs of eigenvalues with positive real part when $A\rightarrow\infty$ (see SI). Strinkingly, both magnitudes coincide, showing that every crossing of eigenvalues takes place from negative to positive real part, consequently the unstable state never recovers its stability and the oscillations are stable for every value of $A>\tilde A$.

Knowing that the system will be oscillating once $A>\ttilde A$ does not give information on the period or amplitude of the oscillations far from $\ttilde A$. One possible approach to studying the frequency of the oscillations far from $\ttilde A$ is to consider the imaginary part of the eigenvalues. If we consider the system with equal degradation rates, the characteristic equation (Equation \ref{eq.characteristic}) corresponds to 
\begin{equation} 
\left( 1+\frac{\lambda}{\delta} \right)^N = -A.
\end{equation}
The eigenvalues are therefore given by $\lambda =\delta(\sqrt[N]{A} \omega_k -1)$ for $k=1,...,N$, with $\omega_k$ the $N$th roots of -1. 
Thus, the eigenvalues with largest real part are $\lambda_1,\lambda_2=\delta[(\sqrt[N]{A}\cos(\pi/N)-1)\pm
i\sqrt[N]{A}\sin(\pi/N)]$. This confirms our result that the steady state is unstable for all $A>\sec^n(\pi/N)$. In addition, perturbing around the steady state, the wavemode
that grows fastest has frequency $\delta\sqrt[N]{A}\sin(\pi/N)$. Although it is tempting to use this value as an approximation for the oscillation frequency far from the Hopf bifurcation, numerical simulations show that this fails to capture the full nonlinear behaviour. Instead they reveal a different scenario in which the oscillations can get slower as the value of $A$ grows (Figure \ref{fig.freqplotsame}b). On the other hand, as expected when moving away from a Hopf bifurcation, the oscillations gain amplitude as $A$ increases (Figure \ref{fig.freqplotsame}b).

\begin{myfloat*}
\begin{mdframed}[backgroundcolor=black!20,linewidth=0.5,leftmargin=-0.2cm,rightmargin=-0.1cm, frametitle={\color{white} Box1. Applications to mRNA and protein dynamics}, frametitlebackgroundcolor=Brown4,linecolor=Brown4
]

So far we have been considered N-component repressive rings without taking any particular consideration of the nature of the biochemical species. It is interesting to focus on the case in which a negative repressive ring includes the mRNA and protein corresponding to each gene as different nodes of the regulatory network. In this scenario, the proteins regulate the mRNA production of other genes, while the mRNA of each gene is translated into the corresponding protein, keeping the same ring topology. Note that since protein translation always increases with the number of mRNA molecules, the number of repressions in the network is the same as for a network where the mRNA is not taken into account. Thus, the same theory developed in the manuscript applies with the difference that the number of nodes $N$ is doubled and two temporal scales for the degradation of mRNA and protein are introduced, the latter being greater than the former. 

One immediate observation is that a two gene negative feedback loop network without mRNA can never oscillate, since the minimum value of $\ttilde A_m$ (Equation \ref{eq.minA}) tends to infinity for $N=2$, making it impossible to find any set of parameters or regulatory functions able to make the system oscillate ($A>\ttilde A_m$). The same can also be seen from the Implicit Equation (Equation \ref{eq.implicit}) where each of the two terms in the sum will always be less than $\pi/2$ for finite positive values of $\alpha$ and $\delta$. By contrast, this is no longer true when mRNA is included in the description, since in this case, $N=4$ and there is a finite value of $\ttilde A_m= 4$, allowing the system to oscillate for values of $A>\ttilde A>\ttilde A_m$, even though the gene network topology is the same. This explains the computational observations of Hazimanikatis and Lee \cite{Hatzimanikatis1999}, in which they study the danger of the common simplification of considering mRNA dynamics to be so fast that they can be considered in equilibrium. Concretely they observe that, for a two-gene feedback loop, considering mRNA to be at equilibrium extinguishes the oscillatory behaviour of the network. 
Not only can our analysis explain this behaviour, but it can also give a measure of the contribution of mRNA degradation to the oscillatory behaviour, indicating that the faster the degradation of the mRNA in comparison with the protein, the smaller will be the mRNA ``angular'' contribution to the Implicit Equation (Equation \ref{eq.implicit}).

In order to understand what happens for a larger number of genes we consider the case where there are $M$ genes composed by $M$ mRNAs with degradation rate $\delta_{mRNA}$ and $M$ proteins with degradation rate $\delta_{Prot}$, forming a negative feedback loop of $N=2M$ nodes in total. The critical value of $A$ is given by Equation \ref{eq.tildeAmoduli},
\begin{equation}
\label{eq.A2}
\ttilde A =  \left(1+\frac{\alpha^2}{\delta_{mRNA}^2} \right)^{M}\left(1+\frac{\alpha^2}{\delta_{Prot}^2} \right)^{M},
\end{equation}
where the value of the angular velocity $\alpha$ is analytically available from the Implicit Equation (Equation \ref{eq.implicit}),
$\tan^{-1}\left(\frac{\alpha}{\delta_{mRNA}}\right)+\tan^{-1}\left(\frac{\alpha}{\delta_{Prot}}\right)=\pi/M$ giving,

\begin{equation}
\label{eq.alpha2}
\alpha= \frac{ -(\delta_{mRNA}+\delta_{Prot})+\sqrt{(\delta_{mRNA}+\delta_{Prot})^2+4\tan^2 (\pi/m) \delta_{Prot} \delta_{mRNA}}}{2 \tan \pi/M}. 
\end{equation}
Substituting Equation \ref{eq.alpha2} into Equation \ref{eq.A2}, it is straightforward to see that the value of $\ttilde A$ is solely determined by the ratio of degradation rates $\delta_{mRNA}/\delta_{Prot}$, and reaches a minimum of $\sec^N(\pi/N)$, as expected, when the two degradation rates coincide (Figure \ref{fig.AmRNA}a).

In the specific case of the two node network of \cite{Hatzimanikatis1999} (\emph{i.e}. $M=2$), we get 
\begin{equation} 
\label{eq.alphamRNA}
\alpha = \sqrt{\delta_{mRNA} \delta_{Prot}} 
\end{equation}
and the critical value of $A$ is 
\begin{equation} 
\label{eq.AmRNA}
\ttilde A = \left( 1 + \frac{\delta_{Prot}}{\delta_{mRNA}} \right) \left( 1+ \frac{\delta_{mRNA}}{\delta_{Prot}} \right) .
\end{equation}
The nice simple form of Equation \ref{eq.alphamRNA} shows that at the bifurcation, oscillations occur on a timescale that depends on both mRNA and protein degradation rates and is intermediate between the two timescales. Additionally as we have already discussed, Equation \ref{eq.AmRNA} implies that $\ttilde A\rightarrow\infty$ when $\delta_{Prot}/\delta_{mRNA}\rightarrow 0$. In particular, the steep variation of $\ttilde A$ with the ratio of the degradation rates makes it very difficult to find an oscillatory network when mRNA has a much faster degradation rate than protein, even when the condition is relaxed and the degradation rate of each of the four species is allowed to vary independently (Figure \ref{fig.ringtrans2}c and \ref{fig.ringtrans2}d ). We see in the figure that the four degradation rates need to be similar in order for oscillations to occur.

\end{mdframed}
\end{myfloat*}

\begin{myfloat*}
\begin{mdframed}[backgroundcolor=black!20,linewidth=0.5,leftmargin=-0.2cm,rightmargin=-0.1cm, frametitle={\color{white} Box1. Applications to mRNA and protein dynamics}, frametitlebackgroundcolor=Brown4,linecolor=Brown4
]

Interestingly this strict condition does not apply to bigger networks. The dependence of $\ttilde A$ on $\delta_{mRNA}/\delta_{Prot}$ becomes less steep as $M$ increases and $\ttilde A$ still takes a finite value of $\sec^M(\pi/M)$ as $\delta_{mRNA}/\delta_{Prot} \rightarrow 0$ (Figure \ref{fig.AmRNA}).\footnote{In the limit as $M \rightarrow \infty$, both $\ttilde A_m$ and the value of $\ttilde A$ as $\delta_{Prot}/\delta_{mRNA} \rightarrow 0$ tend to one, so the dependence on $\delta_{mRNA}/\delta_{Prot}$ disappears.} 
Similar things can be observed if we allow all the degradation rates to be different and we screen numerically for sets of degradation rates that give rise to oscillations (Figure \ref{fig.ringtrans2}c and \ref{fig.ringtrans2}d). We see already for $M=3$ that the actual values of the mRNA degradation rates are relatively unimportant (provided they are constrained to be higher than the protein degradation rates) and the possibility of oscillations is almost exclusively constrained by the degradation rates of the proteins, that are required to be similar.

As in the previous section, is also interesting to study the behaviour of the oscillations far from the Hopf bifurcation when the mRNA is taken into account. As expected from the analysis, the introduction of new species slows down the system, yielding slower oscillations for all the values of $A$ (Figure \ref{freqplot3}b). Additionally, as was shown in the previous section, more species do not necessarily have an effect on the amplitude of the oscillations, which remain the same whether or not mRNA dynamics are considered.

\begin{center}
\includegraphics[width=0.9\columnwidth]{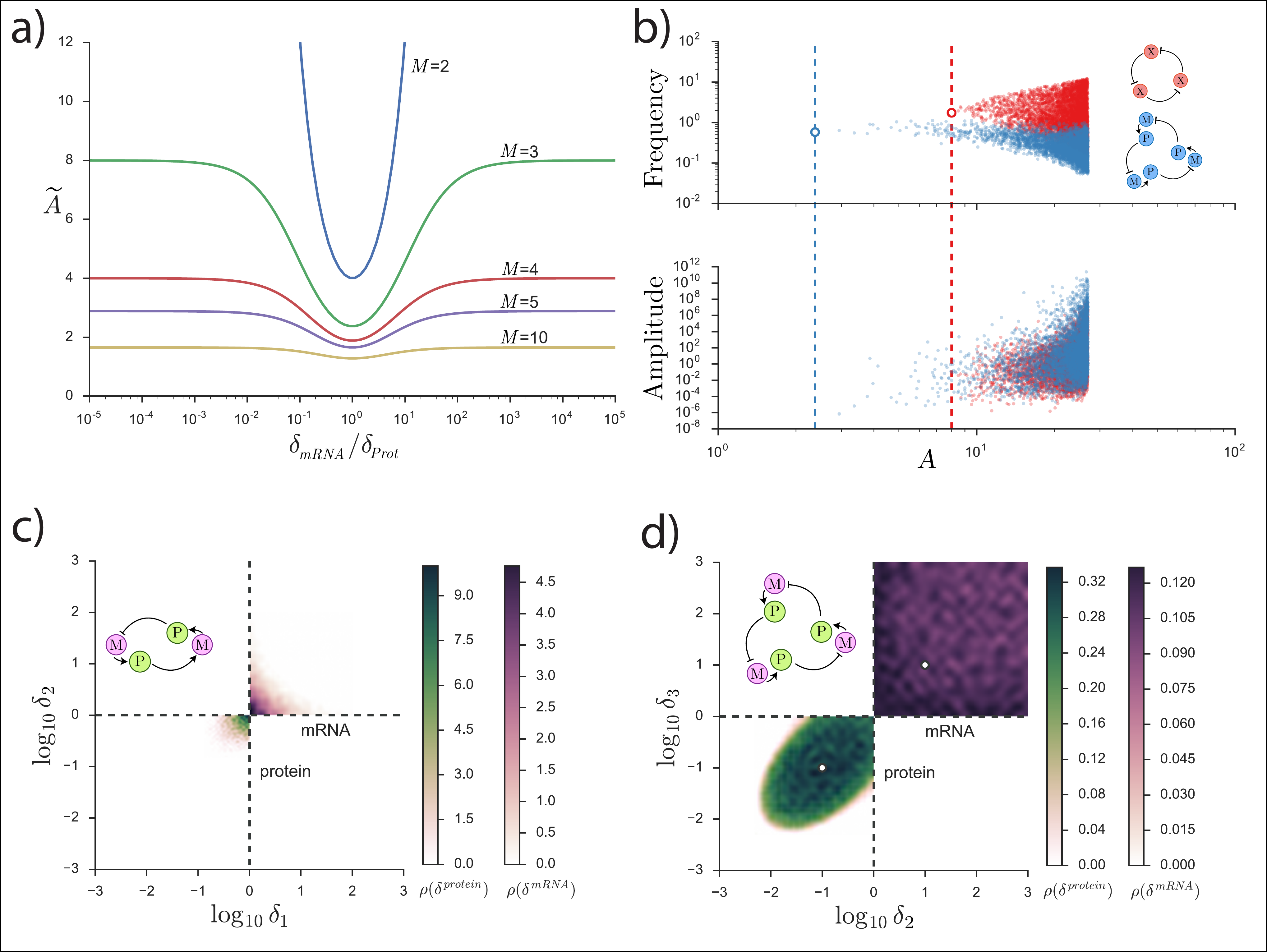}
\captionof{figure}{\label{fig.AmRNA} \label{fig.ringtrans2} \label{freqplot3}  Effects of mRNA and protein degradation time scale differences.  a) Dependence of critical value of $A$ on the ratio of the degradation rates of the protein and mRNA for a system composed of M genes with protein degradation rate $\delta_{Prot}$ and mRNA degradation rate $\delta_{mRNA}$. b) Frequency and amplitude as a function of the network parameter $A$ for 6000 successfully oscillatory networks from a random screening. The random screening was performed for the repressilator network simulated as direct repression between three genes (\emph{red}) and as a six element network (\emph{blue}) taking into account separately mRNA from protein dynamics. Dashed lines and rings show the minimum critical value $\tilde A_m$ and angular velocity at that point $\alpha_m$  predicted by Equations (\ref{eq.minA}) and (\ref{eq.minalpha}) for $N=3$ and $N=6$. Repression functions and parameter screening were the same as in Fig. \ref{fig.freqplotsame}b with additional screening on the degradation rates $\delta_{Prot}: [10^{-3},1]$ and $\delta_{mRNA}: [1,10^3]$, keeping one of the degradation rates fixed as $\delta_{Prot 1}=1$. The translation of mRNA M into protein P is considered to be linear as $f_p(m)=a_p m$, where $a_p$ was also logarithmically sampled ($a_p : [1,10^8]$). c,d) Probability density of oscillations for the two (c)) and three (d)) gene network with mRNA ($M=2$ and $M=3$) for different sets of networks and degradation parameters. The degradation parameters of each test were sampled logarithmically from the ranges $\delta_{mRNA}=[1,10^3]$ (upper quadrant) and $\delta_{Prot}=[10^{-3},1]$ (lower quadrant). Colours show the successfully oscillatory behaviour probability density of a network as a function of pairs of $\delta_{mRNA}$ and $\delta_{Prot}$. For the case $M=3$, one of the species had fixed degradation rates given by $\delta_{mRNA_1}=10$ and $\delta_{Prot_1}=0.1$ (white circles). The random sampling of the other parameters of the network was the same as in Fig. \ref{fig.freqplotsame}b. 
}
\end{center}

\end{mdframed}
\end{myfloat*}

\clearpage
\newpage

\section*{Discussion}

The results obtained in the current study rely on working out properties of the eigenvalues of the system without determining exactly their values. Concretely, we find that the eigenvalues only depend on the values of the degradation rates and a single parameter $A$ that summarizes all the topology, regulatory functions, and specific parameters of the network. The power of this finding is that it allowed us to delve into details of the oscillatory behaviour that are universal for any repressive ring. The main conclusion deriving from this analysis is the requirement of identical degradation rates for all the genes in order to optimize the parameter space that allows oscillations. This property holds independently of how asymmetric the different regulatory functions are. Furthermore, it also yields a quantification of the range of heterogeneity among the degradation rates that can still allow oscillations. This information is valuable from the point of view of synthetic biology where fine tuning of the network is required to optimize oscillatory behaviour. 

The limitations of these findings come in the indetermination of how different regulatory functions affect the actual values of the parameter $A$, suggesting a natural continuation of the research on the topic. Understanding how different biological parameters affect the value of $A$ will lead to knowledge of how these parameters affect the properties of the oscillations of the system and how achievable is the oscillatory condition $A>\ttilde A$. Similarly, we found that the approach fails to predict details of the oscillatory behaviour, such as frequency or amplitude far from the bifurcation point. Results show that for identical topologies, increasing $A$ can lead to increasing or decreasing frequency, suggesting that further knowledge beyond $A$ is required to address these questions.

One of the factors that allowed the analysis was the existence of only one steady state. More complex ring GRNs that include bidirectional interactions can lead to richer bifurcation pictures \cite{Potapov2015}. This hinders the extension of our results to more complex networks. A recent study on the AC-DC network, consisting of a repressilator with an extra cross-repression, is one example of this. Strikingly, for this network, optimization of the oscillatory behaviour revealed that, again, homogeneity of the degradation rates was required. Such results hint at the possibility of extending our current analysis to more complex topologies.   

Finally, the current study was limited to networks that involve a negative feedback loop. Nevertheless, there is also a body of research devoted to understand the oscillations of positive feedback ring GRNs \cite{Smith1987,Mueller2006}. Even though the oscillatory orbits are unstable, they can show long-lived oscillations that allow fast controllable transients between oscillatory and non-oscillatory regimes \cite{Strelkowa2010}. The appeal of such networks, also indicate a possible continuation of our work, seeking to understand the role of degradation rates homogeneity and number of nodes in the nature of such oscillations.

\subsubsection*{Acknowledgements}

RPC and KMP would like to acknowledge support from the Wellcome Trust (grant reference WT098325MA)

\clearpage
\newpage      

\section*{Supplementary Information}

\subsection*{Unattainability of chaotic regimes}
One of the difficulties in extending the result to more than three dimensions is the possibility of chaotic attractors in the system. Nevertheless, monotone systems such as the ones described in equation (\ref{eq.Ndimrep}), follow a Poincar\'{e}-Bendixson-type result \cite{Mallet-Paret1990}. This implies that the $\omega$-limit set of an orbit is either a steady state or a limit cycle. According to theorem $4.1$ of \cite{Mallet-Paret1990}, the Poincar\'{e}-Bendixson-type result applies when $\mathbb{R}_{+}^n$ is positively invariant for monotone cyclic feedback systems defining a negative feedback loop that contain a unique critical point $x^*$ with $\det(-J)>0$. In our system, at the unique fixed point $\det(-J)=(1+A)\prod_{n=1}^N \delta_n>0$ (see eq. (\ref{eq.detJN})). Additionally it is easy to show that  $\mathbb{R}_{+}^n$ is positively invariant since  at each hyperplane $x_n=0$ the dynamics of the $n$-th component follows $\dot x_n=\delta_n f_n(x_{n-1})>0$ (see eq. (\ref{eq.Ndimrep})). As a result of this theorem, the system can only converge to a steady state or a limit cycle. The possible bifurcations in the N-dimensional space have a 2-dimensional analogue, and the change of stability of a system with $\prod_{n=1}^N \lambda _n \neq0$ (see eq. (\ref{eq.detJN})) can only correspond to a Hopf bifurcation. 

\subsection*{Derivation of the Implicit Equation}

Introducing the Hopf bifurcation condition $\lambda = i \alpha$  in Equation \ref{eq.detJN} we obtain,

\begin{equation} 
\label{eq.tildeAN}
\prod_{n=1}^N \left(1+\frac{i\alpha}{\delta_n} \right) =-\ttilde A. 
\end{equation}

Taking the argument of both sides and using the identity $\sum_{i=1}^n \arg(z_i)
\equiv \arg \left ( \prod_{i=1}^n z_i\right)$, we obtain,
\begin{equation}
\label{eq.firstimplicitSI}
\sum_{n=1}^N \tan^{-1} (\alpha/\delta_n) = \pi+2\pi k,\quad k=0,1,2,...
\end{equation}
Similarly, taking the squared moduli of both sides of Equation \ref{eq.tildeAN},
\begin{equation}
\label{eq.tildeAmoduliSI}
\prod_{n=1}^N \left( 1 + \frac{\alpha^2}{\delta_n^2} \right) = {\ttilde A}^2.
\end{equation}
The left hand side is increasing in $\alpha$, so to find the smallest value of $A$ for which there is a pair of imaginary eigenvalues, we must find the smallest $\alpha$ which satisfies Equation \ref{eq.firstimplicitSI}. This will happen when the sum of the arguments equals $\pi$ ($k=0$), because each of the terms in the sum in Equation \ref{eq.firstimplicitSI} runs from $0$ to $\pi/2$ as $\alpha$ increases. This reduces the Implicit Equation for $\alpha$ to,
\begin{equation}
\label{eq.implicitSI}
\sum_{n=1}^N \tan^{-1} (\alpha/\delta_n) = \pi.
\end{equation}

\subsection*{Increase in oscillation probability after introducing a new species}

The change of the parameter size for which the system oscillates as a function of the $N+1$th degradation rate of a ring of $N+1$ species can be obtained from Eq. \ref{eq.tildeAmoduli} as 
\begin{equation}
\frac{\mathrm d \ln \ttilde A}{\mathrm d \delta_{N+1}} = \frac{\alpha^2}{\delta_{N+1}(\delta_{N+1}^2+\alpha^2)} \left[ \frac{\sum_{n=1}^{N+1} \frac{\delta_{N+1}}{\delta_n^2+\alpha^2}}{\sum_{n=1}^{N+1} \frac{\delta_{n}}{\delta_n^2+\alpha^2}} -1 \right],
\end{equation}
which means that $\frac{\mathrm d \ttilde A}{\mathrm d \delta_{N+1}}>0$ if and only if $\sum_{n=1}^N \sin^2 \theta_n (\cot \theta_{N+1}-\cot \theta_{n})>0$. Since increasing $\delta_{N+1}$, while keeping the other degradation rates fixed, decreases $\theta_{N+1}$ and increases the other $\theta_n$, every term in this sum is increasing in $\delta_{N+1}$. This means that $\ttilde A$ decreases with $\delta_{N+1}$ until it reaches a certain value (which must be within the range of values of the other degradation rates). Above this value $\ttilde A$ increases. Thus the probability of oscillations increases from zero when $\delta_{N+1}$ is zero to a maximum value for some optimal value of $\delta_{N+1}$ (intermediate between the $\delta_n$, $n=1,\hdots, N$) and then decreases to the same probability as for the network without species $N+1$ as $\delta_{N+1}$ tends to infinity.

\subsection*{Persistence of instability for $A>\tilde A$}

The characteristic polynomial (Equation \ref{eq.characteristic}) is analytic in $\lambda$ and $A$. Therefore by the implicit function theorem, the roots of the equation are analytic in $A$, except at the turning points of the polynomial. At these turning points it is straightforward to check that two negative real eigenvalues collide to form a pair of complex conjugate eigenvalues in such a way that the real and imaginary parts on each branch are continuous in $A$. The other eigenvalues are analytic in $A$ at these points.
We have already shown in the previous section that there are no zero eigenvalues for any value of $A$, therefore the real parts of an eigenvalue can only change sign as a pair of eigenvalues crossing the imaginary axis. From the multiplicity of the Implicit Equation (Equation \ref{eq.firstimplicitSI}), we already saw that there will be other values of $\alpha_k>\alpha$ for which a pair of eigenvalues crosses the imaginary axis. In our analysis in the previous section we chose $\alpha$ to give the minimum value of $A=\ttilde A$. The other points where eigenvalues cross the imaginary axis will occur at $\ttilde A_k > \ttilde A$. This is clear from the LHS of Equations \ref{eq.firstimplicitSI} and \ref{eq.tildeAmoduli}, which are both increasing functions of $\alpha$. 

The RHS of the Implicit Equation (Equation \ref{eq.firstimplicitSI}) has infinitely many possible values, i.e. $\sum_{n=1}^N \tan^{-1} (\alpha/\delta_{n})= \pi, 3 \pi, 5 \pi,\dots$. Nevertheless, each of the terms of the sum can contribute at most $\pi/2$ to the result. Therefore, the sum has the upper bound $\sum_{n=1}^N \tan^{-1} (\alpha/\delta_n) \le N\pi/2$, and for finite $\alpha$ (necessary since $A$ is finite), the inequality is strict. Therefore, when $N=3,4,5,6$ there can only be one possible solution $\alpha$ of the Implicit Equation; when $N=7,8,9,10$ there can be two values of $\alpha$, etc. Concretely, the maximum number of times there are eigenvalues crossing the imaginary axis is $\lfloor (N+1)/4 \rfloor$, where $\lfloor \cdot \rfloor$ is the \emph{floor} operator.

Each time a crossing occurs, it must only consist of a single pair of eigenvalues. This can be seen by supposing the opposite case in which $i \alpha$ is a repeated root of the characteristic equation (Equation \ref{eq.characteristic}), $P(i\alpha) = \prod_{n=1}^N (1+\frac{i\alpha}{\delta_n}) + A= 0$. In this case it will also be a zero of the derivative of the characteristic function $P'(i\alpha) = 0 = \sum_{n=1}^N (P(i\alpha) - A)/   (\delta_n+ i\alpha) = -A \sum_{n=1}^N\frac{\delta_n-i\alpha}{{\delta_n}^2+\alpha^2}$, which is impossible since each term in the final expression has negative real part.

On the other hand, as $A \rightarrow \infty$, the characteristic equation (Equation \ref{eq.characteristic}) requires the moduli of the roots to be arbitrarily large, so that the value of $\lambda/\delta_n$ will eventually greatly outweigh $1$ and $\lambda$ will approach a value satisfying,
\begin{equation}
\lambda ^N = -A \prod_{n=1}^n\delta_n.
\end{equation}
This has solutions $\lambda_k = \sqrt[N]{A \prod_{n=1}^N\delta_n} \omega_k$ with $k=1,...,N$, where $\omega_k=\mathrm e^{(2k-1)\pi i/N}$ are each of the $N$ solutions to the $N$th root of $-1$.  Therefore, the number of eigenvalues with a positive real part will be the number of $\omega_k$ with $\arg(\omega_k) \in (-\pi/2,\pi/2)$, which for $N>2$ is 2$\lfloor (N+1)/4 \rfloor$. Thus, there are twice as many roots with positive real part as there are pairs of eigenvalues crossing the imaginary axis as $A$ goes from $0$ to $\infty$.  This shows that every pair of eigenvalues crossing the imaginary axis must do so by changing the real part from negative to positive as $A$ increases. Therefore, the steady state is unstable for all $A>\tilde A$. \\

\subsection*{Simulations of networks}

The networks were simulated using Python custom code integrated using the PyDSTools dynamical systems environment  \cite{Clewley2012}. Trajectories were solved by precompiling in C a multistep Radau method. Bifurcation diagrams were obtained using continuation techniques and integrations under the same Python environment. 
 
\onecolumn

\subsection*{First Lyapunov coefficient of the Hopf bifurcation}

In the system with the same repressive function, $f$, for each species, the steady state $F(x^*)=x^*$ occurs when $f(x^*)=x^*$. A Hopf bifurcation occurs at $\tilde x$, where $f(\tilde x)=\tilde x$ and $f'(\tilde x)=-\sec(\pi/N)$.

\vspace{4mm}

In order to study the nature of the Hopf bifurcation, it is necessary to compute the First Lyapunov coefficient $ \ell_1$ evaluated at the critical point $\tilde x$, that can be obtained as \cite{Kuznetsov2013},
\begin{eqnarray}
\label{eq.l1}
\ell_1(\tilde x)&=&\frac{1}{2\alpha}\Re\!\left[\langle p,\!C(q,q,\bar q)\rangle\!-\!2\langle p, B(q, J^{-1}B(q,\bar q))\rangle+\right.\nonumber\\
&& \left.\langle p,B(\bar q,(2i\alpha I_N-J)^{-1}B(q,q))\rangle \right],\nonumber\\
&&
\end{eqnarray}
which involves the computation of inner products of vectors in $\mathbb{C}^N$: $\langle a,b\rangle=\bar a^T b$. The matrix $J$ in Equation \ref{eq.l1} is the Jacobian of the dynamical system at the critical point. For the system with same repressive functions and degradation rates for each species this is given by
\[
J =\left( \begin{array}{ccccc} -1 & 0 & \ldots & 0 & -\sec(\pi/N) \\ -\sec(\pi/N) & -1 & 0 & \ldots & 0 \\ 0 & \sec(\pi/N) & -1 & 0 & \ldots \\ \ddots \\ \ 0 & \ldots & 0 & -\sec(\pi/N) & -1 \end{array} \right). 
\]
Note that for the sake of simplicity we have scaled the time with the degradation rate $\delta$. Additionally, $\alpha=\tan(\pi/N)$ in Equation \ref{eq.l1} is the modulus of the pair of eigenvalues at the Hopf bifurcation and $q$ the corresponding eigenvector
\[ q = \left( \begin{array}{c} 1/\sqrt{N} \\ \omega/\sqrt{N} \\ \vdots \\ \omega^{N-1}/\sqrt{N} \end{array} \right), \] where $\omega=-\cos(\pi/N)+i\sin(\pi/N)$ is an $N$th root of unity. On the other hand, $p$ is the eigenvalue of $J^T$, corresponding to eigenvalue $-i\tan(\pi/N)$, that in this case is the same, $p=q$.
The only missing ingredients to start computing $\ell_1(\tilde x)$ are the definition of the bilinear and trilinear forms
\[
B_j(y,z) = \sum_{k,l=1}^N \left. \frac{\partial^2f_j}{\partial x_kx_l} \right|_{\tilde x} y_kz_l, \quad
\]
\[
C_j(y,z,w) = \sum_{k,l=1}^N \left. \frac{\partial^3f_j}{\partial x_kx_lx_m} \right|_{\tilde x} y_kz_lw_m \quad \forall j=1,...,N.
\]
Now we can proceed by calculating sequentially the different terms of Equation \ref{eq.l1}. The evaluation of the bilinear forms $B(q,q)$ and $B(q,\bar q)$ is immediate,
\[
B(q,q) = f''(\tilde x) \left( \begin{array}{c} \omega^{2N-2}/N \\ 1/N \\ \omega^2/N \\ \vdots \\ \omega^{2(N-2)}/N \end{array} \right), \quad
 B(q,\bar{q}) =  \frac{f''(\tilde x)}{N} \left( \begin{array}{c} 1 \\ 1 \\  \vdots \\ 1 \end{array} \right). 
 \]
 Thus, the product $J^{-1}B(q,\bar q)$ requires the knowledge of the row sums of $J^{-1}$. Since the row sums of $J$ are the same and equal to $(-1-sec(\pi/N))$, $J$ is a stochastic matrix multiplied by $-1-sec(\pi/N)$. Thus $J^{-1}$ will also be a stochastic matrix multiplied by $(-1-\sec(\pi/N))^{-1}$, and its product with the bilinear form will be, 
\[ J^{-1} B(q,\bar{q}) = \frac{-f''(\tilde x)}{N(1+\sec(\pi/N))} \left( \begin{array}{c} 1 \\ 1 \\  \vdots \\ 1 \end{array} \right). \]
So
\[ B(q,J^{-1}B(q,\bar{q})) = f''(\tilde x)  \frac{-f''(\tilde x)}{N(1+\sec(\pi/N))} \left( \begin{array}{c} \omega^{N-1}/\sqrt{N} \\ 1 \\ \omega\sqrt{N} \\ \vdots \\ \omega^{N-2}/\sqrt{N} \end{array} \right), \]
and hence
\begin{eqnarray}
\bar{p}^T  B(q,J^{-1}B(q,\bar{q})) &=& \frac{-f''(\tilde x)^2}{N^2(1+\sec(\pi/N))} [N\omega^{N-1}]\nonumber \\
&\hspace{-6cm}=& \hspace{-3cm}\frac{f''(\tilde x)^2}{N(1+\sec(\pi/N))} [\cos(\pi/N)+i\sin(\pi/N)]. 
\end{eqnarray}
Next, in order to compute the third term of Equation \ref{eq.l1} we need to work with the inverse $(2\tan(\pi/N)i-J)^{-1}$, which has rows which are permutations of each other as,
\[ (2\tan(\pi/N)i-J)^{-1} = \left( \begin{array}{cccc} b_1 & b_2 & \ldots & b_N \\ b_N & b_1 & \ldots & b_{N-1} \\ b_{N-1} & b_1 & \ldots & b_{N-2} \\ \ddots &&& \\ b_2 & b_3 & \ldots & b_1 \end{array} 
\right). \]
So 
\begin{eqnarray*} 
&&\hspace{-1cm}(2\tan(\pi/N)i-J)^{-1} B(q,q) =\\
&&\hspace{-1cm}=\frac{f''(\tilde x)}{N} \left( \begin{array}{c} b_1 \omega^{2(N-1)} + b_2 + b_3 \omega^2+\ldots+ b_N \omega^{2(N-2)} \\ b_N \omega^{2(N-1)}+ b_1 + b_2 \omega^2 + \ldots + b_{N-1} \omega^{2(N-2)} \\ \vdots \\ b_2 \omega^{2(N-1)} + b_3 + \ldots + b_N \omega^{2(N-3)} + b_1 \omega^{2(N-2)} \end{array} \right) \\ 
&&\hspace{-1cm}= \frac{f''(\tilde x)}{N} [b_2 \omega^{2(N-1)} + b_3 + \ldots + b_N \omega^{2(N-3)} + b_1 \omega^{2(N-2)}] \left( \begin{array}{c} \omega^2 \\ \omega^4 \\ \ldots \\ \omega^{2(N-1)} \\ 1 \end{array} \right) \\
 &&\hspace{-1cm}= [b_1 + b_2 \omega^2 + \ldots + b_N \omega^{2(N-1)}] B(q,q). 
\end{eqnarray*}

This means that $B(q,q)$ is an eigenvector of $(2\tan(\pi/N)i-J)^{-1}$ with eigenvalue $[b_1 + b_2 \omega^2 + \ldots + b_N \omega^{2(N-1)}]$. Thus it must also be an eigenvector of $(2\tan(\pi/N)i-J)$ with eigenvalue $\frac{1}{b_1 + b_2 \omega^2 + \ldots + b_N \omega^{2(N-1)}}$. On the other hand we know that,
\[ (2\tan(\pi/N)i-J) B(q,q) = \frac{f''(\tilde x)}{N} \left( \begin{array}{c} (2i\tan(\pi/N)+1) \omega^{2N-2} +  \sec(\pi/N) \omega^{2(N-2)} \\ \vdots \end{array} \right), \]
which means that the eigenvalue is 
\[ 1+2i\tan(\pi/N)+ \sec(\pi/N)\bar{\omega}^2, \]
therefore
\[ B(\bar{q},(2\tan(\pi/N)i-J)^{-1} B(q,q)) = f''\frac{f''(\tilde x)}{N\sqrt{N}} [b_1 + b_2 \omega^2 + \ldots + b_N \omega^{2(N-1)}] 
\left( \begin{array}{c} \bar{\omega}^3 \\ \bar{\omega}^2 \\ \bar{\omega} \\ 1 \\ \omega \\ \vdots  \end{array} \right). \]
And the third inner product of Equation \ref{eq.l1} can be written as, 
\begin{eqnarray*} \bar{p}^T  B(\bar{q},(2\tan(\pi/N)i-J)^{-1} B(q,q)) & = & \frac{f''(\tilde x)^2}{N^2}[b_2 \omega^{2(N-1)} + b_3 + \ldots + b_N \omega^{2(N-3)} + b_1 \omega^{2(N-2)}] N \bar{\omega}^3 \\ & = & \frac{f''(\tilde x)^2 \bar{\omega}^3}{N}\frac{1}{1+2i\tan(\pi/N)+ \sec(\pi/N)\bar{\omega}^2,}\end{eqnarray*}
which has a real part,
\[\Re{\left( \bar{p}^T  B(\bar{q},(2\tan(\pi/N)i-J)^{-1} B(q,q))  \right)} = \frac{f''(\tilde x)^2}{N} \frac{c(2c-1)(2c^2-c-2)}{(1+c)(5-4c)},\] where $c = \cos(\pi/N)$.\\
Finally, we need to use the trilinear form to compute the first term of Equation \ref{eq.l1},
\[ C(q,q,\bar{q} ) = \frac{f'''(\tilde x)}{N\sqrt{N}} \left( \begin{array}{c} \omega^{N-1} \\ 1 \\ \omega \\ \vdots \\ \omega^{N-2} \end{array} \right)  \]
and so
\[ \bar{p}^T C(q,q,\bar{q}) =  \frac{f'''(\tilde x)}{N^2} N \bar{\omega} = \frac{f'''(\tilde x)}{N} \bar{\omega}. \]
Putting the three terms together, the first Lyapunov coefficient is given by (with $c\equiv\cos(\pi/N)$):
\begin{eqnarray} l_1(\tilde x) & = & \frac{1}{2\tan(\pi/N)} \Re \left[ \frac{f'''(\tilde x)}{N} \bar{\omega} + 2  \frac{f''(\tilde x)^2}{N(1+\sec(\pi/N))} \bar{\omega} +\frac{f''(\tilde x)^2}{N} \frac{c(2c-1)(2c^2-c-2)}{(1+c)(5-4c)}\right] \nonumber \\ & = &  \frac{1}{2N\tan(\pi/N)} \left[-cf'''(\tilde x) +\frac{-2c+\frac{(2c-1)(2c^2-c-2)}{(5-4c)}}{1+\sec(\pi/N)}f''(\tilde x)^2 \right]  \nonumber \\ & = &  \frac{c^2}{2N\sin(\pi/N)}  \left[ -f'''(\tilde x)+\frac{4c^3+4c^2-13c+2}{(1+c)(5-4c)}f''(\tilde x)^2 \right].\end{eqnarray} 
This quantity does not have a determined sign and therefore the Hopf bifurcation can be supercritical or subcritical depending on the details of $f(x)$. For instance, for the repressive function $f(x)=a/(1+(1+x-x^2+x^3)^h)$ the sign of $\ell_1$ will depend on the coefficient $h$ (see Fig.\ref{fig.lyapunov}).

The Hopf bifurcation is supercritical if and only if
\begin{equation} \label{Supercrit}-f'''(\tilde x)+\frac{4c^3+4c^2-13c+2}{(1+c)(5-4c)}f''(\tilde x)^2 < 0 .\end{equation}
The coefficient multiplying $f''(\tilde x)^2$ is negative for all $N$ and decreases monotonically from $-2/3$ for $N=3$ to $-3/2$ as $N \rightarrow \infty$. This means that the criterion on $f$ at the fixed point necessary for supercriticality becomes less stringent as $N$ increases, so the Hopf bifurcation (for a given $f$) is more likely to be supercritical as $N$ increases. 

If, as in \cite{Buse2009}, we consider $f$ to be a repressive Hill function, \emph{i.e.} $f(x)=c/(1+x^r)$, then 
\[ \frac{-f'''(\tilde x)}{f''(\tilde x)^2} = \frac{r^2-15r+38}{2(r-5)^2} .\]
For $r >2$, this is less than $2/3$ and therefore the Hopf bifurcation is supercritical for any $N \ge 3$. Note that we have shown that the Hopf bifurcation occurs when 
$f'(\tilde x)=-\sec (\pi/N)$, but for the Hill function system $f'(\tilde x)=-r\tilde x^r/(1+\tilde x^r)$, so $|f'(\tilde x)|<r$, so for the Hopf bifurcation to occur requires $r>\sec(\pi/N) \ge 2$. Therefore the Hopf bifurcation only occurs for $r>\sec(\pi/N)\ge 2$ and is always supercritical. This extends the result of \cite{Buse2009} to the case of $N$ species.

\clearpage
\renewcommand\thefigure{S.\arabic{figure}}
\setcounter{figure}{0}
\begin{figure}[h]
\centering
\includegraphics[width=0.7\columnwidth]{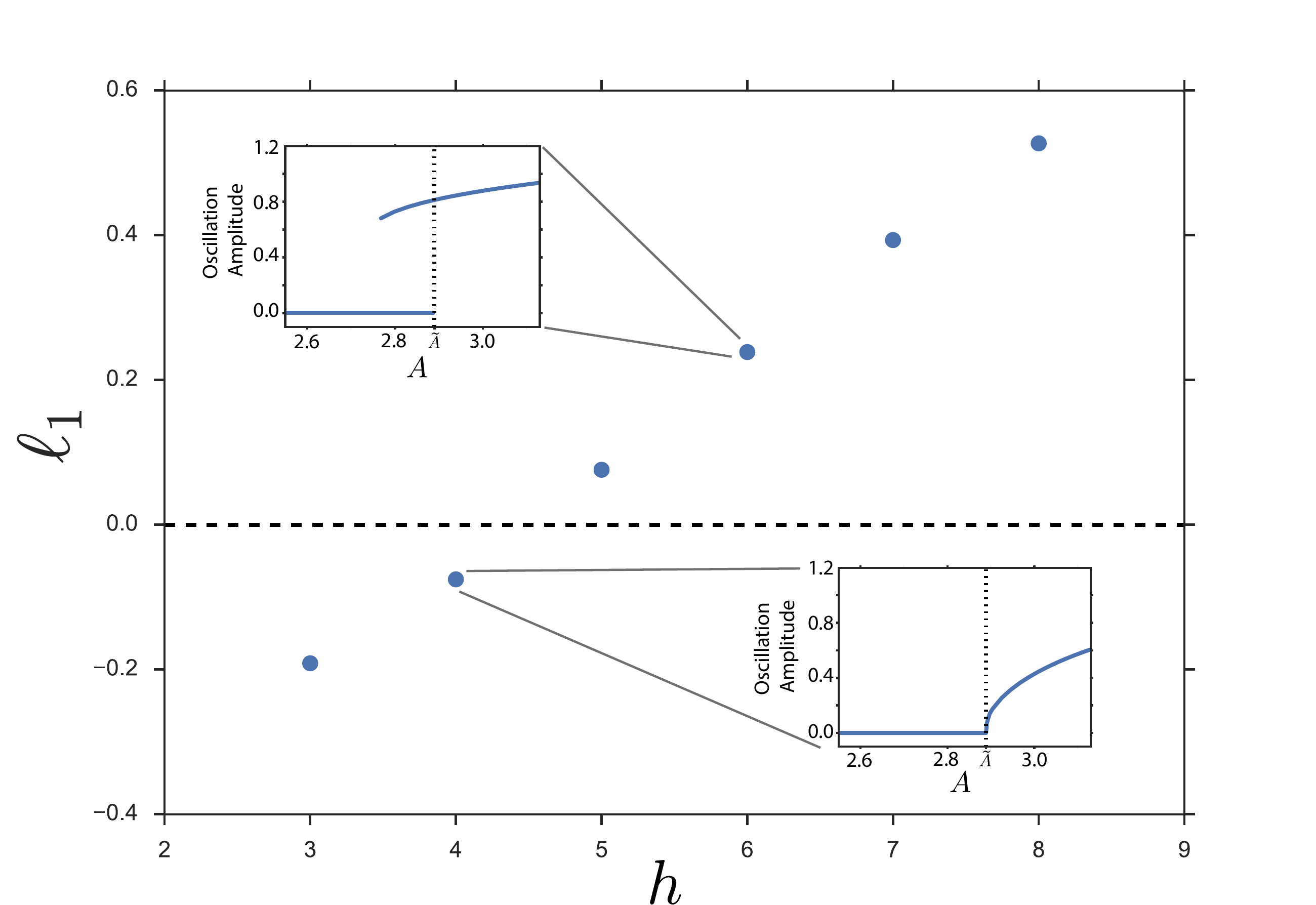}
\caption{\label{fig.lyapunov} Depedence of sign of the first Lyapunov coefficient $\ell_1$ with the exponent $h$ for the repressive regulatory function $f(x)=a/(1+(1+x-x^2+x^3)^h)$ for a ring of 5 genes with identical degradation rates. Inset) Bifurcation diagrams for different values of $\ell_1$ show how its sign determines if the Hopf bifurcation is supercritical or subcritical.}
\end{figure}


\begin{thebibliography}{}

\bibitem[Bachmair {\rm et~al.}, 1986]{Bachmair1986}
Bachmair, A., Finley, D.  and Varshavsky, A. (1986{\rm{}}).
\newblock In vivo half-life of a protein is a function of its amino-terminal
  residue.
\newblock {\rm Science } \emph{234}, 179--186.

\bibitem[Bar-Or {\rm et~al.}, 2000]{Bar-Or2000}
Bar-Or, R.~L., Maya, R., Segel, L.~A., Alon, U., Levine, A.~J.  and Oren, M.
  (2000{\rm{}}).
\newblock Generation of oscillations by the p53-Mdm2 feedback loop: a
  theoretical and experimental study.
\newblock {\rm Proceedings of the National Academy of Sciences } \emph{97},
  11250--11255.

\bibitem[Belle {\rm et~al.}, 2006]{Belle2006}
Belle, A., Tanay, A., Bitincka, L., Shamir, R.  and O'Shea, E.~K.
  (2006{\rm{}}).
\newblock {Quantification of protein half-lives in the budding yeast proteome}.
\newblock {\rm Proc. Natl. Acad. Sci.} \emph{103}, 13004--13009.

\bibitem[Bu\c{s}e {\rm et~al.}, 2009]{Buse2009}
Bu\c{s}e, O., Kuznetsov, A.  and P{\'e}rez, R.~A. (2009{\rm{}}).
\newblock Existence of limit cycles in the repressilator equations.
\newblock {\rm International Journal of Bifurcation and Chaos } \emph{19},
  4097--4106.

\bibitem[Bu\c{s}e {\rm et~al.}, 2010]{Buse2010}
Bu\c{s}e, O., P{\'e}rez, R.  and Kuznetsov, A. (2010{\rm{}}).
\newblock Dynamical properties of the repressilator model.
\newblock {\rm Physical Review E } \emph{81}, 066206.

\bibitem[Christiano {\rm et~al.}, 2014]{Christiano2014}
Christiano, R., Nagaraj, N., Fr{\"{o}}hlich, F.  and Walther, T.~C.
  (2014{\rm{}}).
\newblock {Global Proteome Turnover Analyses of the Yeasts S. cerevisiae and S.
  pombe}.
\newblock {\rm Cell Rep.} \emph{9}, 1959--1965.

\bibitem[Clewley, 2012]{Clewley2012}
Clewley, R. (2012{\rm{}}).
\newblock {Hybrid Models and Biological Model Reduction with PyDSTool}.
\newblock {\rm PLoS Comput. Biol.} \emph{8}, e1002628.

\bibitem[Davidson and Levine, 2005]{Davidson2005}
Davidson, Eric, H. and Levine, M. (2005{\rm{}}).
\newblock {Gene regulatory networks}.
\newblock {\rm Proceedings of the National Academy of Sciences } \emph{102},
  4935.

\bibitem[Dequ{\'e}ant {\rm et~al.}, 2006]{Dequant2006}
Dequ{\'e}ant, M.-L., Glynn, E., Gaudenz, K., Wahl, M., Chen, J., Mushegian, A.
  and Pourqui{\'e}, O. (2006{\rm{}}).
\newblock A complex oscillating network of signaling genes underlies the mouse
  segmentation clock.
\newblock {\rm Science } \emph{314}, 1595--1598.

\bibitem[Dice, 1987]{Dice1987}
Dice, J. (1987{\rm{}}).
\newblock Molecular determinants of protein half-lives in eukaryotic cells.
\newblock {\rm The FASEB Journal } \emph{1}, 349--357.

\bibitem[Elowitz and Leibler, 2000]{Elowitz2000}
Elowitz, M.~B. and Leibler, S. (2000{\rm{}}).
\newblock {A synthetic oscillatory network of transcriptional regulators.}
\newblock {\rm Nature } \emph{403}, 335--338.

\bibitem[Estrada {\rm et~al.}, 2016]{Estrada2016}
Estrada, J., Wong, F., DePace, A.  and Gunawardena, J. (2016{\rm{}}).
\newblock {Information Integration and Energy Expenditure in Gene Regulation}.
\newblock {\rm Cell } \emph{166}, 234--244.

\bibitem[Fraser and Tiwari, 1974]{Fraser1974}
Fraser, A. and Tiwari, J. (1974{\rm{}}).
\newblock Genetical feedback-repression: II. Cyclic genetic systems.
\newblock {\rm Journal of theoretical biology } \emph{47}, 397--412.

\bibitem[Garcia-Ojalvo {\rm et~al.}, 2004]{Garcia-Ojalvo2004}
Garcia-Ojalvo, J., Elowitz, M.~B.  and Strogatz, S.~H. (2004{\rm{}}).
\newblock Modeling a synthetic multicellular clock: repressilators coupled by
  quorum sensing.
\newblock {\rm Proceedings of the National Academy of Sciences of the United
  States of America } \emph{101}, 10955--10960.

\bibitem[Hatzimanikatis and Lee, 1999]{Hatzimanikatis1999}
Hatzimanikatis, V. and Lee, K.~H. (1999{\rm{}}).
\newblock {Dynamical Analysis of Gene Networks Requires Both mRNA and Protein
  Expression Information}.
\newblock {\rm Metabolic Engineering } \emph{1}, E1--E7.

\bibitem[Hirata {\rm et~al.}, 2002]{Hirata2002}
Hirata, H., Yoshiura, S., Ohtsuka, T., Bessho, Y., Harada, T., Yoshikawa, K.
  and Kageyama, R. (2002{\rm{}}).
\newblock Oscillatory expression of the bHLH factor Hes1 regulated by a
  negative feedback loop.
\newblock {\rm Science } \emph{298}, 840--843.

\bibitem[Hoffmann {\rm et~al.}, 2002]{Hoffmann2002}
Hoffmann, A., Levchenko, A., Scott, M.~L.  and Baltimore, D. (2002{\rm{}}).
\newblock The I$\kappa$B-NF-$\kappa$B signaling module: temporal control and
  selective gene activation.
\newblock {\rm Science } \emph{298}, 1241--1245.

\bibitem[K. {\rm et~al.}, 2015]{Parmar2015}
K., P., K.B., B., Y.N., K.  and S.J., H. (2015{\rm{}}).
\newblock {Time-Delayed Models of Gene Regulatory Networks}.
\newblock {\rm Computational and Mathematical Methods in Medicine }
  \emph{2015}.

\bibitem[Kuznetsov, 2013]{Kuznetsov2013}
Kuznetsov, Y.~A. (2013{\rm{}}).
\newblock {\rm Elements of applied bifurcation theory}, vol. 112,.
\newblock Springer Science \& Business Media.

\bibitem[Levine and Davidson, 2005]{Levine2005}
Levine, M. and Davidson, E.~H. (2005{\rm{}}).
\newblock {Gene regulatory networks for development}.
\newblock {\rm Proceedings of the National Academy of Sciences } \emph{102},
  4936--4942.

\bibitem[Mallet-Paret and Smith, 1990]{Mallet-Paret1990}
Mallet-Paret, J. and Smith, H. (1990{\rm{}}).
\newblock {The Poincare Bendixson Theorem for Monotone Cyclic Feedback
  Systems}.
\newblock {\rm Journal of Dynamics and Differential Equations } \emph{2},
  367--421.

\bibitem[Michael and Oren, 2003]{Michael2003}
Michael, D. and Oren, M. (2003{\rm{}}).
\newblock The p53--Mdm2 module and the ubiquitin system.
\newblock {\rm Seminars in cancer biology } \emph{13}, 49--58.

\bibitem[Monk, 2003]{Monk2003}
Monk, N.~A. (2003{\rm{}}).
\newblock Oscillatory expression of Hes1, p53, and NF-$\kappa$B driven by
  transcriptional time delays.
\newblock {\rm Current Biology } \emph{13}, 1409--1413.

\bibitem[M{\"u}ller {\rm et~al.}, 2006]{Mueller2006}
M{\"u}ller, S., Hofbauer, J., Endler, L., Flamm, C., Widder, S.  and Schuster,
  P. (2006{\rm{}}).
\newblock A generalized model of the repressilator.
\newblock {\rm Journal of mathematical biology } \emph{53}, 905--937.

\bibitem[Niederholtmeyer {\rm et~al.}, 2015]{Niederholtmeyer2015}
Niederholtmeyer, H., Sun, Z., Hori, Y., Yeung, E., Verpoorte, A., Murray, R.~M.
   and Maerkl, S.~J. (2015{\rm{}}).
\newblock Rapid cell-free forward engineering of novel genetic ring
  oscillators.
\newblock {\rm Elife } \emph{4}, e09771.

\bibitem[Olson, 2006]{Olson2006}
Olson, E.~N. (2006{\rm{}}).
\newblock {Gene regulatory networks in the evolution and development of the
  heart}.
\newblock {\rm Science } \emph{313}, 1922--1927.

\bibitem[Panovska-Griffiths {\rm et~al.}, 2013]{Panovska-Griffiths2013}
Panovska-Griffiths, J., Page, K.~M.  and Briscoe, J. (2013{\rm{}}).
\newblock {A gene regulatory motif that generates oscillatory or multiway
  switch outputs.}
\newblock {\rm Journal of the Royal Society, Interface / the Royal Society }
  \emph{10}, 20120826.

\bibitem[Pigolotti {\rm et~al.}, 2007]{Pigolotti2007}
Pigolotti, S., Krishna, S.  and Jensen, M.~H. (2007{\rm{}}).
\newblock Oscillation patterns in negative feedback loops.
\newblock {\rm Proceedings of the National Academy of Sciences } \emph{104},
  6533--6537.

\bibitem[Potapov {\rm et~al.}, 2015]{Potapov2015}
Potapov, I., Zhurov, B.  and Volkov, E. (2015{\rm{}}).
\newblock {Multi-stable dynamics of the non-adiabatic repressilator}.
\newblock {\rm Journal of The Royal Society Interface } \emph{12},
  20141315--20141315.

\bibitem[Purcell {\rm et~al.}, 2010]{Purcell2010}
Purcell, O., Savery, N.~J., Grierson, C.~S.  and di~Bernardo, M. (2010{\rm{}}).
\newblock {A comparative analysis of synthetic genetic oscillators.}
\newblock {\rm Journal of the Royal Society Interface } \emph{7}, 1503--1524.

\bibitem[Reddy and Rey, 2014]{Reddy2014}
Reddy, A.~B. and Rey, G. (2014{\rm{}}).
\newblock Metabolic and nontranscriptional circadian clocks: eukaryotes.
\newblock {\rm Annual review of biochemistry } \emph{83}, 165--189.

\bibitem[Sauka-Spengler and Bronner-Fraser, 2008]{Sauka-Spengler2008}
Sauka-Spengler, T. and Bronner-Fraser, M. (2008{\rm{}}).
\newblock A gene regulatory network orchestrates neural crest formation.
\newblock {\rm Nature reviews Molecular cell biology } \emph{9}, 557--568.

\bibitem[Smith, 1987]{Smith1987}
Smith, H. (1987{\rm{}}).
\newblock Oscillations and multiple steady states in a cyclic gene model with
  repression.
\newblock {\rm J. Math. Biol } \emph{25}, 169--190.

\bibitem[Strelkowa and Barahona, 2010]{Strelkowa2010}
Strelkowa, N. and Barahona, M. (2010{\rm{}}).
\newblock {Switchable genetic oscillator operating in quasi-stable mode}.
\newblock {\rm Journal of The Royal Society Interface } \emph{7}, 1071--1082.

\bibitem[Stricker {\rm et~al.}, 2008]{Stricker2008}
Stricker, J., Cookson, S., Bennett, M.~R., Mather, W.~H., Tsimring, L.~S.  and
  Hasty, J. (2008{\rm{}}).
\newblock {A fast, robust and tunable synthetic gene oscillator}.
\newblock {\rm Nature } \emph{456}, 516--519.

\bibitem[Tuttle {\rm et~al.}, 2005]{Tuttle2005}
Tuttle, L.~M., Salis, H., Tomshine, J.  and Kaznessis, Y.~N. (2005{\rm{}}).
\newblock Model-driven designs of an oscillating gene network.
\newblock {\rm Biophysical journal } \emph{89}, 3873--3883.

\end{thebibliography}
\end{document}